\documentclass[fleqn,usenatbib]{mnras}

\usepackage[T1]{fontenc}
\DeclareRobustCommand{\VAN}[3]{#2}
\let\VANthebibliography\thebibliography
\def\thebibliography{\DeclareRobustCommand{\VAN}[3]{##3}\VANthebibliography}

\usepackage{graphicx}	% Including figure files
\usepackage{amsmath}	% Advanced maths commands
\usepackage{amssymb}	% Extra maths symbols

\title[Evidence for powerful winds]{Evidence for powerful winds and the
associated reverse shock as the origin of the Fermi bubbles}

\author[Yutaka Fujita]{
Yutaka Fujita$^{1}$\thanks{E-mail: y-fujita@tmu.ac.jp}
\\
$^{1}$Department of Physics, Graduate School of Science, Tokyo Metropolitan University, \\
1-1 Minami-Osawa, Hachioji-shi, Tokyo 192-0397, Japan
}

\date{Accepted XXX. Received YYY; in original form ZZZ}

\pubyear{2022}

\begin{document}
\label{firstpage}
\pagerange{\pageref{firstpage}--\pageref{lastpage}}
\maketitle

\begin{abstract}
The Fermi bubbles are large gamma-ray-emitting structures. They are
symmetric about the Galactic Centre (GC), and their creation is
therefore attributed to intensive energy injection at the GC. In this
study, we focus on the non-equilibrium X-ray gas structures associated
with the bubbles. We show that a combination of the density,
temperature, and shock age profiles of the X-ray gas can be used to
distinguish the energy injection mechanisms. By comparing the results of
numerical simulations with observations, we indicate that the bubbles
were created by a fast wind from the GC because it generates a strong
reverse shock and reproduces the observed temperature peak there. On the
other hand, instantaneous energy injection at the GC cannot reproduce
the temperature profile. The wind had a speed of $\sim 1000\rm\: km\:
s^{-1}$, and blew for $\sim 10^7$~yr. Because the mass flux of the wind
is large, the entrainment of interstellar gas by wide-angle outflows
from the black hole is required. Thus, the wind may be the same as
active galactic nuclei outflows often observed in other galaxies and
thought to regulate the growth of galaxies and their central black
holes.
\end{abstract}


%%%%%%%%%%%%%%%%%%%%%%%%%%%%%%%%%%%%%%%%%%%%%%%%%%

%%%%%%%%%%%%%%%%% BODY OF PAPER %%%%%%%%%%%%%%%%%%

\section{Introduction}

The Fermi bubbles are large gamma-ray-emitting structures extending
$\sim \pm 50^\circ$ above and below the Galactic plane
\citep{2010ApJ...717..825D,2010ApJ...724.1044S,2014ApJ...793...64A}, and related structures
have been discovered based on X-ray \citep{2018MNRAS.480..223K} and
radio observations
\citep{2004ApJ...614..186F,2008ApJ...680.1222D,2013A&A...554A.139P}.
However, the origin of the Fermi bubbles, for example, the instantaneous
explosive activity of a central supermassive black hole (Sagittarius
A$^*$;
\citealp{2012ApJ...756..181G,2013ApJ...775L..20F,2014ApJ...789...67F,2020ApJ...894..117Z,2022MNRAS.514.2581M,2022NatAs...6..584Y}),
winds from the black hole
\citep{2012MNRAS.424..666Z,2014ApJ...790..109M,2015ApJ...811...37M}, a
starburst
\citep{2013Natur.493...66C,2014MNRAS.444L..39L,2015MNRAS.453.3827S}, or
steady star-formation activity
(\citealt{2011PhRvL.106j1102C,2012MNRAS.423.3512C}, see also \citealt{2017MNRAS.468.3051N}), has been debated.

Peculiar X-ray structures were recognised in the direction of the
Galactic Centre (GC) even before the discovery of Fermi bubbles in gamma
rays. In particular, the regions corresponding to high X-ray and radio
wave intensities are called the North Polar Spur and Loop~I
\citep{1995ApJ...454..643S}. There has been discussion on whether their
origin is from a nearby supernova remnant or galactic-scale structures
located near the GC \citep{1983BASI...11....1S,2000ApJ...540..224S,2003ApJ...582..246B}.
However, recent eROSITA telescope observations have shown that these
X-ray structures are part of a larger structure encompassing the Fermi
bubbles, which means that their association with the galactic structure
and Fermi bubbles has been confirmed \citep{2020Natur.588..227P}.
Consequently, the properties of the X-ray structures are related to the
formation process of the Fermi bubbles. Previous studies have mainly
focused on the morphology of X-ray structures that reflect gas density
distributions \citep[e.g.][]{2022NatAs...6..584Y}. By
contrast, in this study, we demonstrate that the combined analysis of
the density, temperature, and shock age profiles of the gases based on
X-rays is a powerful tool for distinguishing the energy-injection
mechanisms of the Fermi bubbles. Importantly, this helps us understand
the origin of the Fermi bubbles. Here, the shock age is defined as the
time elapsed since the passing of a shock wave through a gas element.

The paper is organised as follows. In Section~\ref{sec:model}, we
explain our models and numerical setup. In section~\ref{sec:result},
we compare our simulation results with observations, and show that the
Fermi bubbles should have been created by a fast wind. In
section~\ref{sec:discuss}, we discuss the origin of the wind based on
the parameters derived in Section~\ref{sec:result}. The conclusion of
this paper is presented in Section~\ref{sec:conc}.

\section{Models}
\label{sec:model}

To study the dependence of the X-ray gas profiles on different
energy-injection mechanisms at the GC, we performed one-dimensional (1D)
Lagrangian hydrodynamic simulations of shock propagation from the GC
\citep{2007nmai.conf.....B}, regarding the Fermi bubbles as a single
bubble centred on the GC. The one-dimensionality of the code
enables unknown parameters to be specified by performing $\sim 10^4$
simulations and comparing the results with observations. Because the
actual Fermi bubbles are not spherically symmetric, we adopted the
volume-averaged radius ($r\sim 4$~kpc) as the typical radius of the
bubbles \citep{2015ApJ...808..107C}. Radiative cooling is ignored
because the densities of the ejecta and halo gas are low. The code
allows us to explicitly determine the shock age by measuring the time
elapsed since shock passage at each gas element. The position of the
shock wave can be identified by searching for the position where the
flow converges.

\subsection{Galaxy model}

The gravitational potential is given by
\begin{equation}
 \Phi = \Phi_{\rm halo} + \Phi_{\rm disk} + \Phi_{\rm bulge}\:,
\end{equation}
where
\begin{equation}
 \Phi_{\rm halo}(r) = v_{\rm halo}^2\ln(r^2 + d_h^2)
\end{equation}
is the halo potential,
\begin{equation}
 \Phi_{\rm disk}(r) = -\frac{G M_{\rm disk}}{a+\sqrt{r^2 + b^2}}
\end{equation}
is the potential known as the Miyamoto-Nagai disk, $G$ 
is the gravitational constant \citep{1975PASJ...27..533M}, and
\begin{equation}
 \Phi_{\rm bulge}(r) = -\frac{G M_{\rm bulge}}{r+d_b}
\end{equation}
is the potential, known as the Hernquist stellar bulge. The adopted
values are $v_{\rm halo}=131.5\:\rm km\: s^{-1}$, $d_h=12$~kpc, $M_{\rm
disk}=10^{11}\: M_\odot$, $a=6.5$~kpc, $b=0.26$~kpc, $M_{\rm
bulge}=3.4\times 10^{10}\: M_\odot$, and $d_b=0.7$~kpc
\citep{2012ApJ...761..185Y,2014ApJ...789...67F}. The influence of
gravity was minor because supersonic flows were considered.

\subsection{Energy injection}

\begin{figure}
	\includegraphics[width=\columnwidth]{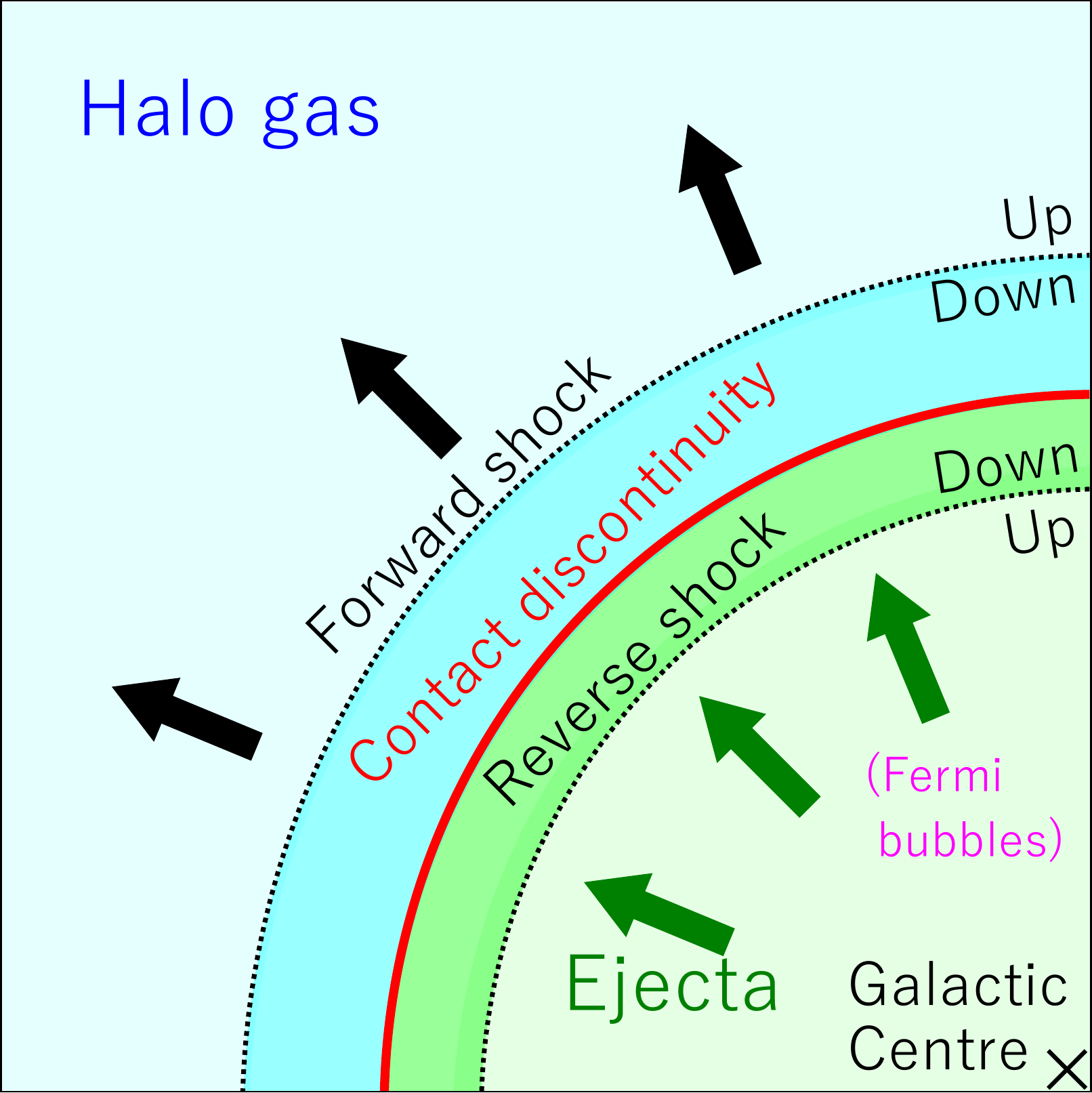}
    \caption{Schematic of the structures surrounding the Fermi bubbles. Cyan and green areas are filled with halo gas and ejecta (wind gas), respectively. The contact discontinuity (red solid line) is the boundary between the two gases. The forward and reverse shocks are shown by the black dotted lines; the upstream side and the downstream side of each shock are indicated by “Up” and “Down,” respectively. Regions heated by the forward and reverse shocks are shown in dark colours. The gamma-ray emitting Fermi bubbles correspond to the region inside the reverse shock (light green).}
    \label{fig:sch}
\end{figure}

We considered two energy-injection types. One is the wind type, in which
a constant wind blows from the GC for a period of $0 < t < t_0$, where
$t=t_0$ is the current time. Specifically, the wind blows into the
galactic halo, and two shock waves are formed. One is the forward shock,
which propagates in the galactic halo gas, and the other is the reverse
shock, which propagates in the wind gas (Fig.~\ref{fig:sch}). In this
context, the wind gas can be regarded as ejecta from the GC. A contact
discontinuity or boundary between the halo gas and ejecta exists between
the two shocks. As for the wind type, we considered two models. One is
the wind-K model, in which we assumed that the wind is cold at the
boundary. Another is the wind-T model, in which the wind gas has a large
thermal energy

The other energy-injection type is the explosion type, for which energy
is instantaneously injected into the GC at $t=0$ as a single
explosion. We also consider two models for this type. In the explosion-K
model, the energy is given as the kinetic energy of the ejecta. Another
one is the explosion-T model, in which the explosion energy is given as
the thermal energy of the ejecta. As in the wind type, the forward shock
is formed in the halo gas, the reverse shock is formed in the ejecta,
and there is contact discontinuity between the two shocks.

At $t=0$, the halo gas is in hydrostatic equilibrium and is isothermal
($T=0.2$~keV; \citealp{2009PASJ...61..805Y}). We set the inner boundary
at $r_{\rm in} =1$ (wind models) or 0.5~kpc (explosion models), and the
outer boundary at $r_{\rm out}=15$~kpc from the GC. The results were
insensitive to the boundary positions when the wind or explosion
conditions were the same. The initial halo gas density at $r=r_{\rm in}$
is treated as a parameter ($\rho_1$).  In the wind models, we injected
the wind (ejecta) gas at $r=r_{\rm in}$ with velocity $u_{\rm w1}$ and
density $\rho_{\rm w1}$. In the wind-K model, the thermal energy of the
wind at $r=r_{\rm in}$ is assumed to be negligible. In the wind-T
model, the wind gas has thermal energy of $3.1$~keV per particle
at $r=r_{\rm in}$. The value of 3.1 keV does not have a solid
physical meaning. We adopted it because it is much larger than the
temperature of the gas around the Galactic centre ($\sim 0.7$~keV;
e.g. \citealt{2019ApJ...875...32N}) and that of the halo gas
($T=0.2$~keV). Thus, the flow of 3.1~keV should be regarded as an
extreme case of hot outflows, and the actual flow is likely to have a
much lower thermal energy.

In the explosion-K model, the energy of the ejecta ($E_{\rm exp}$) is
given in the form of kinetic energy. The ejecta of a mass of $M_{\rm
ej}$ with kinetic energy of $E_{\rm exp}$ is uniformly distributed at
$r_{\rm in}=0.5 < r < 1$~kpc at $t=0$, The velocity of the ejecta is
given such that it increases linearly at $0.5 < r < 1$~kpc. The initial
thermal energy of the ejecta was zero.  In the explosion-T model, the
ejecta of a mass of $M_{\rm ej}$ with thermal energy of $E_{\rm exp}$ is
uniformly distributed at $0.5 < r < 1$~kpc at $t=0$, and the ejecta is
initially at rest.

\section{Results}
\label{sec:result}

\begin{table*}
\begin{minipage}{\textwidth}
\begin{center}
%	\centering
	\caption{Best-fit parameters.}
	\label{tab1}
	\begin{tabular}{lccccccc} % four columns, alignment for each
		\hline
Models & $t_0$ & $u_{\rm w1}$ & $\rho_{\rm w1}/m_{\rm p}$\footnotemark[1] & $M_{\rm ej}$ & $E_{\rm exp}$  & $\rho_1/m_{\rm p}$\footnotemark[1] & $\chi^2$/dof \\
 & (yr) & ($\rm km\: s^{-1}$) & ($\rm cm^{-3}$) & ($M_\odot$) & (erg) & ($\rm cm^{-3}$) &  \\
\hline
wind-K & $9.0\times 10^6$ & $1.1\times 10^3$ & $9.3\times 10^{-3}$ & ... & ... & 0.024 & 51.7/56 \\
wind-T & $9.1\times 10^6$ & $1.7\times 10^2$ & $5.3\times 10^{-2}$ & ... & ... & 0.017 & 55.8/55 \\
explosion-K & $6.5\times 10^6$ & ... & ... & $3.9\times 10^7$ & $3.0\times 10^{56}$ & 0.027 & 190.0/54 \\
explosion-T & $5.7\times 10^6$ & ... & ... & $4.2\times 10^7$ & $6.7\times 10^{56}$ & 0.032 & 242.3/56 \\
		\hline
	\end{tabular}
\end{center}
\footnotemark[1]{$m_{\rm p}$ is the proton mass.}
\end{minipage}
\end{table*}

\begin{figure*}
  \begin{minipage}[b]{0.45\linewidth}
    \centering
    \includegraphics[keepaspectratio, scale=0.55]{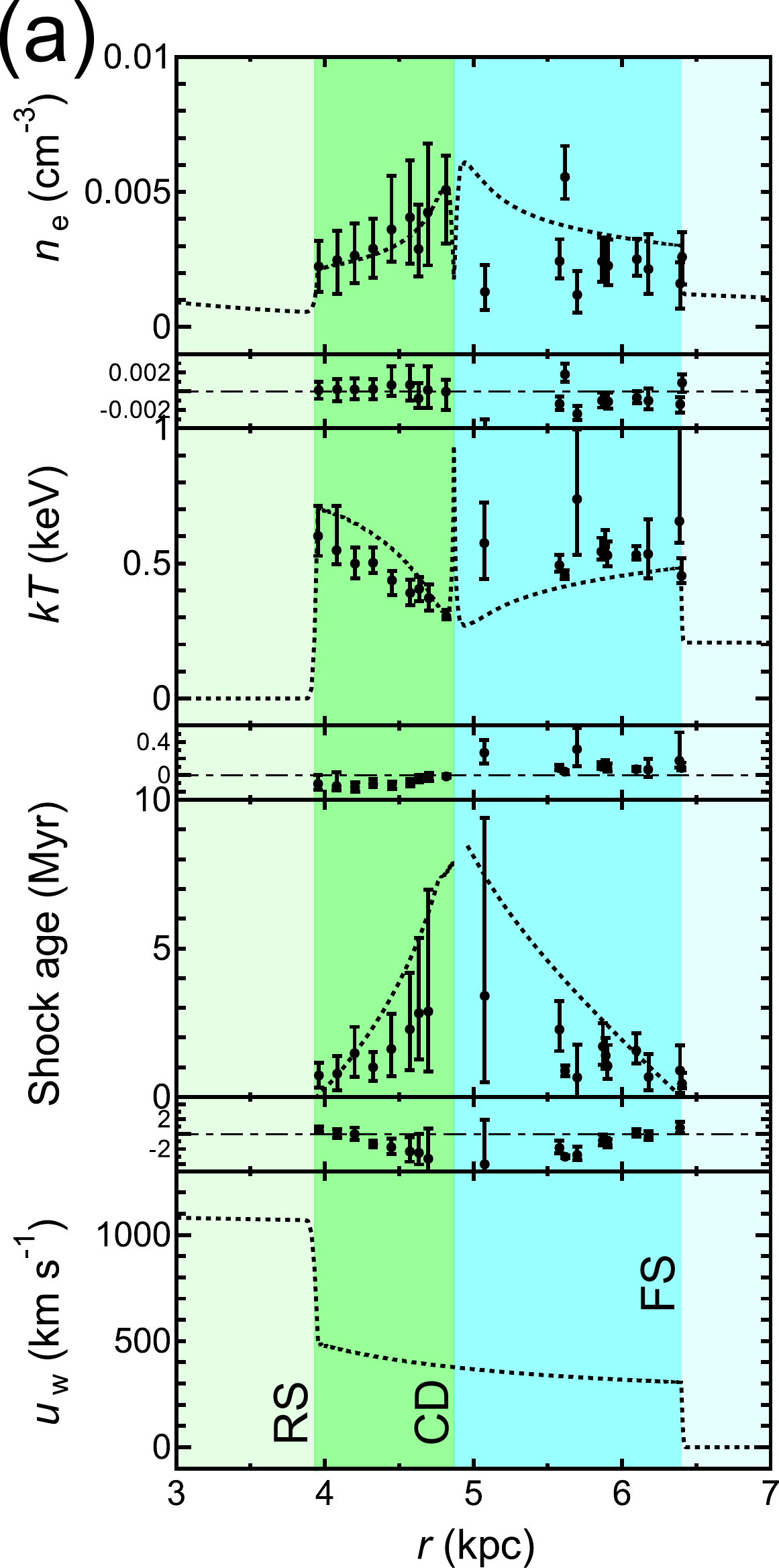}
  \end{minipage}
  \begin{minipage}[b]{0.45\linewidth}
    \centering
    \includegraphics[keepaspectratio, scale=0.55]{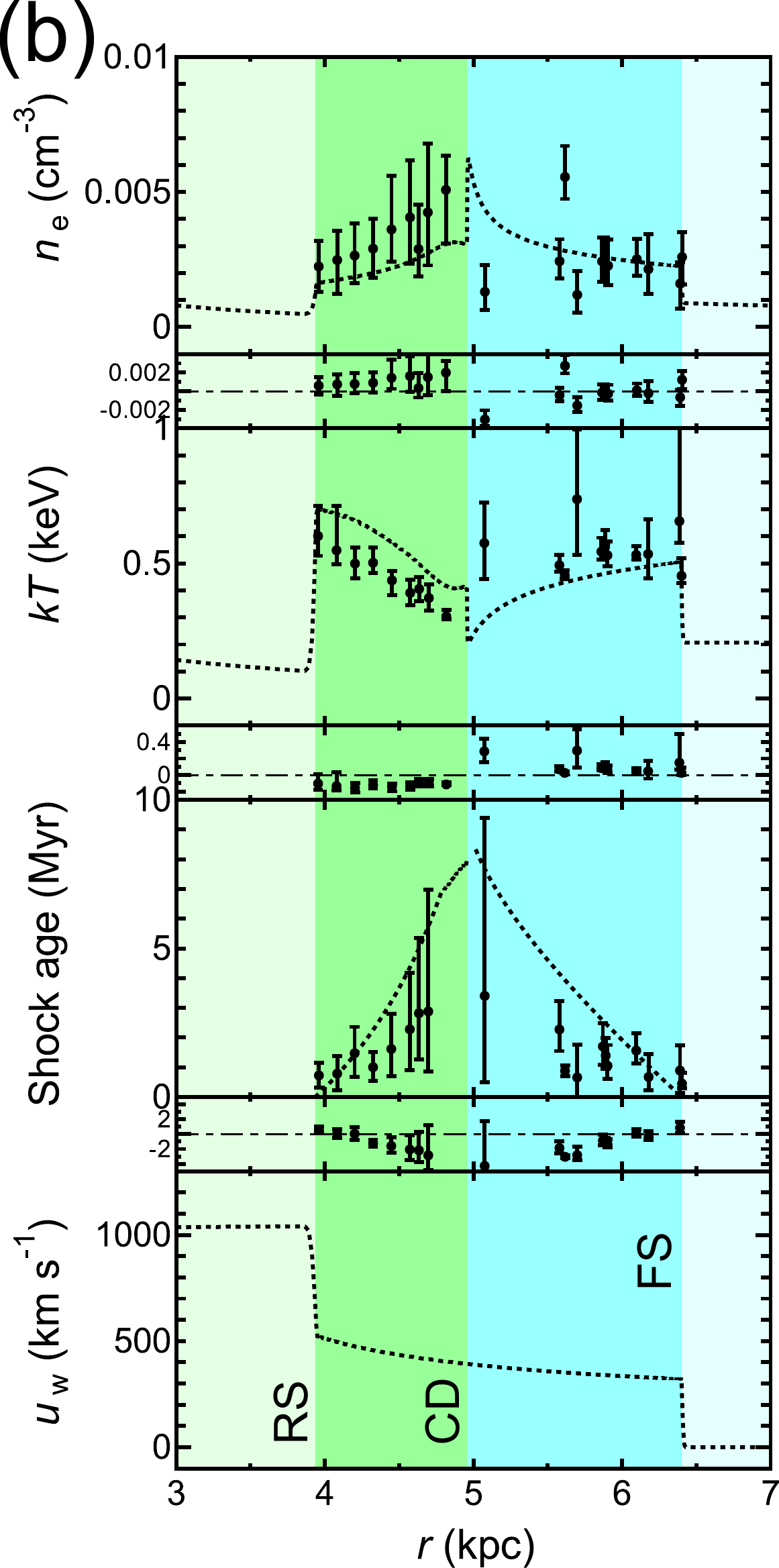}
  \end{minipage}
\caption{The gas profiles for (a) wind-K model, and (b) wind-T
model. The panels are electron density (top), temperature (upper
middle), shock age (lower middle), and velocity (bottom) at $t=t_0$
(current time). The simulation results are shown by the dotted lines,
and the observations are shown by the dots
\citep{2022MNRAS.512.2034Y}. For the latter, the errors are $1\:\sigma$,
and the angular size is transformed into the radius, assuming that the
distance to the GC is 8~kpc. Residual plots are given underneath
each of the panels. The colours correspond to those in
Fig.~\ref{fig:sch}. The boundaries are the reverse shock (RS), the
contact discontinuity (CD), and the forward shock (FS). }
\label{fig:wind}
\end{figure*}

\begin{figure*}
  \begin{minipage}[b]{0.45\linewidth}
    \centering
    \includegraphics[keepaspectratio, scale=0.55]{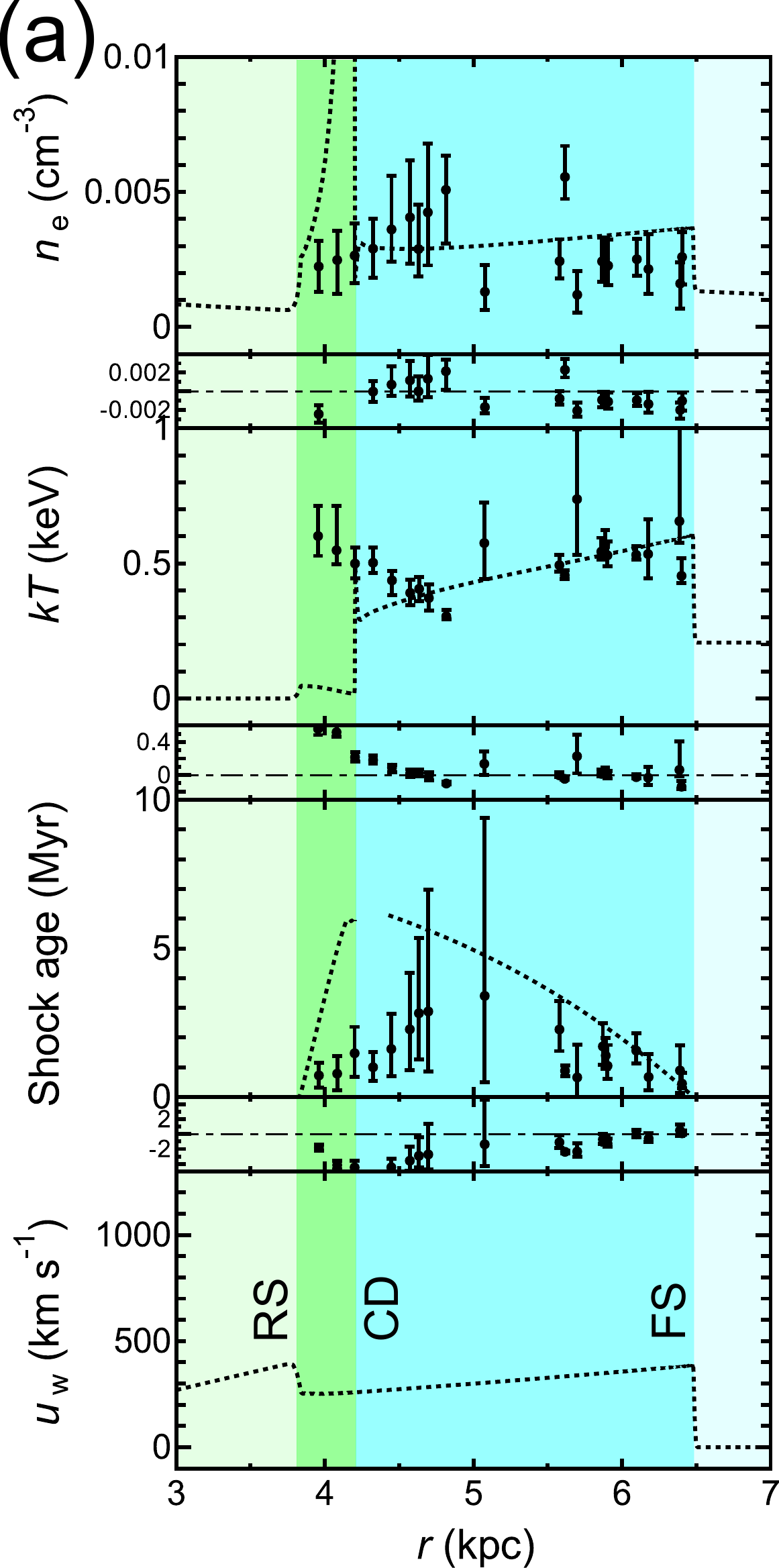}
  \end{minipage}
  \begin{minipage}[b]{0.45\linewidth}
    \centering
    \includegraphics[keepaspectratio, scale=0.55]{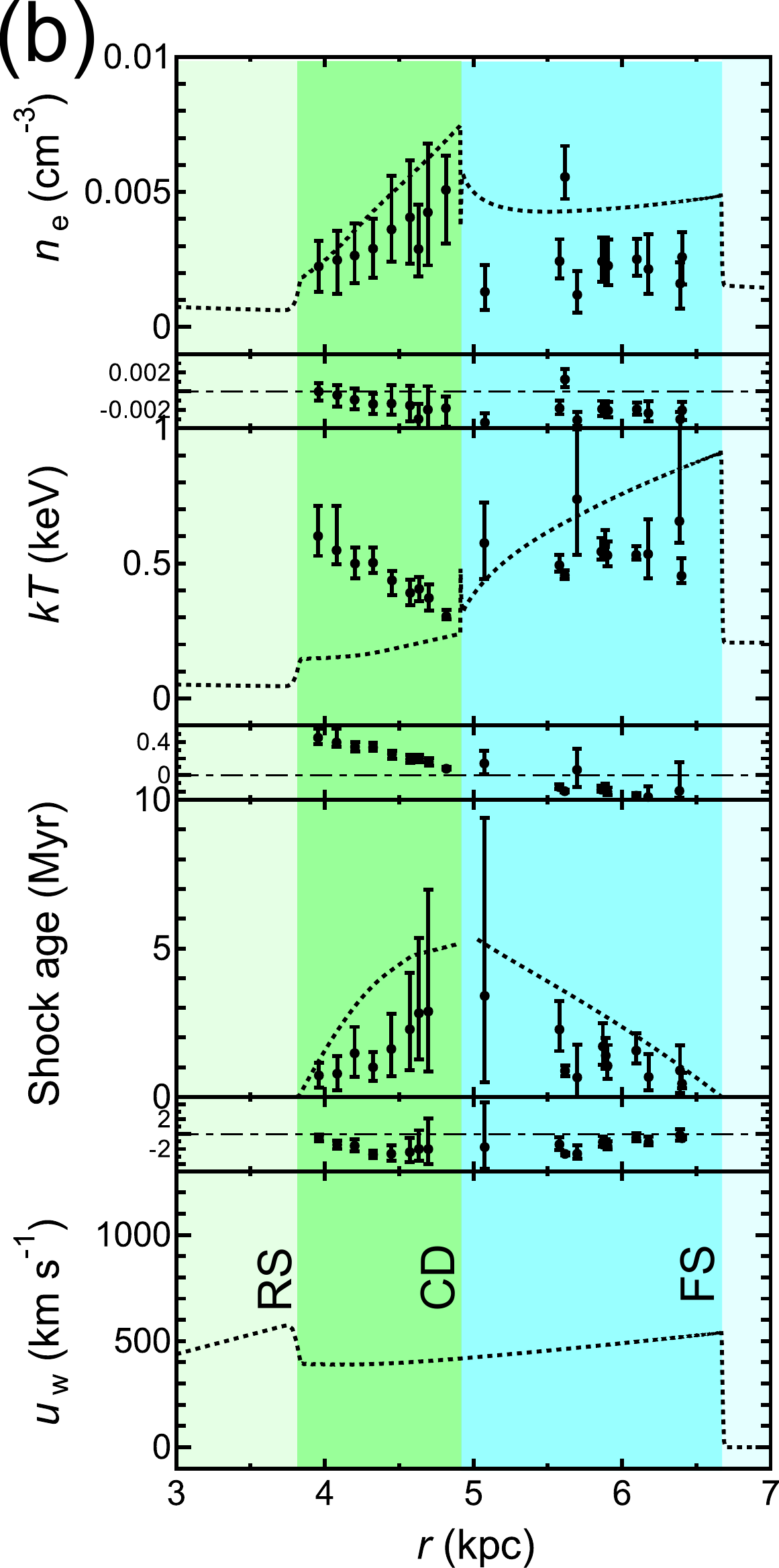}
  \end{minipage}
\caption{Same as Fig.~\ref{fig:wind} but for (a)
explosion-K model, and (b) explosion-T model.} \label{fig:exp}
\end{figure*}

%\subsection{Comparison with observations}

We compared the simulation results with the observed profiles recently
obtained by reanalysing the data from the Suzaku satellite, which has a
high sensitivity to diffuse X-ray emissions \citep{2022MNRAS.512.2034Y}.
In previous X-ray spectral analysis
\citep[e.g.][]{2013ApJ...779...57K,2020ApJ...904...54L}, the collisional
equilibrium (CIE) state was assumed and a plasma model (\texttt{apec})
representing the CIE was used. However, the CIE is not guaranteed for
the Fermi bubbles because the gas density is low and collisions between
the gas particles are not frequent. In particular, the ionisation
temperature at a shock may not be equal to either the kinetic
temperature of the electrons or that of protons\footnote{On the
downstream side of the shocks we considered, we assume that the electron
temperature has well approached the proton temperature through Coulomb
interaction and other mechanisms (e.g. waves excited by plasma
instabilities; \citealt{2020ApJ...904...12B}).}, because the heating of
the plasma occurs in a very short timescale compared to the relaxation
time. Therefore, in their reanalysis, \citet{2022MNRAS.512.2034Y}
adopted a plasma model (\texttt{nei}) representing the non-equilibrium
ionisation (NEI) state and found that \texttt{nei} actually reproduces
the observed spectra better than \texttt{apec}. The X-ray spectrum of
the NEI plasma is different from that of the CIE plasma, and the degree
of the difference depends on the time lapsed since the plasma is forced
out of equilibrium. The \texttt{nei} model can evaluate this time, and
it can be interpreted as the shock age. This approach has often been
adopted for supernova remnants \citep[e.g.][]{2001ApJ...561..308S} and
clusters of galaxies \citep[e.g.][]{2008PASJ...60.1133F}. 

\citet{2022MNRAS.512.2034Y} derived the profiles of the density $n_{\rm
e}$, temperature $kT$, and shock age for a part of the X-ray bright
regions surrounding the Fermi bubbles (North Polar Spur/Loop I). The
observed temperature increases toward the inner ($r\sim 4$~kpc) and
outer ($r\sim 6.5$~kpc) edges of the structure, while the density and
shock age peak around the middle (Figs.~\ref{fig:wind}
and~\ref{fig:exp}). Here, we assumed that the distance to the GC is
8~kpc.

To best reproduce the observed profiles (density, temperature, and shock
age), four simulation parameters were adjusted. For the wind models, the
parameters were the current time $t_0$, wind velocity ($u_{\rm w1}$),
wind density ($\rho_{\rm w1}$), and initial density of the halo gas
($\rho_1$) at $r=1$~kpc from the GC. The explosion model includes $t_0$,
total ejecta mass ($M_{\rm ej}$), total explosion energy ($E_{\rm
exp}$), and $\rho_1$. We performed $\sim 10^4$ simulations for each
model to determine the parameters that reproduced the observed profiles.
We choose only the simulation results in which the reverse shock is at
$r=4.0\pm 0.2$~kpc, and the forward shock is at $r=6.5\pm 0.2$~kpc at
$t=t_0$.  We needed these constraints because the data of density,
temperature, and shock age are not available on the upstream sides of
both reverse and forward shocks ($r \lesssim 4.0$~kpc and $r \gtrsim
6.5$~kpc) due to low X-ray brightness. Without these data, the density
and temperature jumps across the shocks cannot be identified in the
fittings. On the other hand, the positions of the shocks are obvious
from ROSAT X-ray images \citep{2022MNRAS.512.2034Y}.

We simultaneously fitted the simulated profiles to the observed
profiles. The fits were performed in logarithmic parameter spaces. The
best-fit parameters, the associated $\chi^2$ values, and the degree of
freedom (dof) are shown in Table~\ref{tab1}.  The simulation results
with the best fit parameters are presented in Figs~\ref{fig:wind}
and~\ref{fig:exp}. Although the density and shock age profiles are
similar among the two wind and explosion-T models, the temperature and
velocity profiles are intrinsically different. While the wind models
reproduce all observed density, temperature, and shock age profiles, the
explosion models does not reproduce the temperature profile.  In the
explosion-K model (Fig.~\ref{fig:exp}a), the sharp density peak between
the reverse shock and contact discontinuity is inconsistent with the
observations. The peak height of the density is $\sim 0.06\rm\:
cm^{-3}$.

In the wind models (Fig.~\ref{fig:wind}), the forward and reverse shocks
initially formed at the contact discontinuity, where the shock age is
the largest. Subsequently, the gas on the downstream sides of both
shocks was considerably heated. Currently, the inner temperature peak
($r\sim 4$~kpc) corresponds to the reverse shock, and the outer peak
($r\sim 6.4$~kpc) corresponds to the forward shock. In the wind-K model,
the reverse shock is powered by a fast wind blowing from the GC. The
wind velocity is almost constant on the upstream side of the revised
shock ($u_w\sim 1100\rm\: km\: s^{-1}$ at $r< 4$~kpc; bottom of
Fig.~\ref{fig:wind}a). This large wind velocity results in a drastic
velocity decrease to $u_w\sim 500\rm\: km\: s^{-1}$ at the reverse
shock, thereby converting the large kinetic energy into thermal energy
and creating a sharp temperature peak. The contact discontinuity between
the wind gas (ejecta) and halo gas is at the peak of the density and
shock age profiles and at the trough of the temperature profile. 
We note that the sharp temperature peak at $r\sim 4.8$~kpc in
Fig.~\ref{fig:wind}a is a numerical artifact at the contact
discontinuity. The wind-T model shows a similar behaviour
(Fig.~\ref{fig:wind}b). In this model, although the wind velocity at
$r=r_{\rm in}$~kpc ($u_{\rm w1}$) is only $170\rm\: km\: s^{-1}$
(Table~\ref{tab1}), it increased to $\sim 1000\rm\: km\: s^{-1}$
slightly upstream of the reverse shock ($r\lesssim 4$~kpc;
Fig.~\ref{fig:wind}b), which is comparable to that in the wind-K model
(Fig.~\ref{fig:wind}a). By contrast, for the explosion models
(Fig.~\ref{fig:exp}), a prominent temperature peak is not observed at
the reverse shock ($r\sim 3.8$~kpc), although a minor temperature
increase exists there. According to this model, the ejecta expanded
after the explosion. However, the expansion rate has decreased
substantially over the last $t_0\sim 6\times 10^6$~yr
(Table~\ref{tab1}), and the velocity peak at the reverse shock is at
most comparable to that at the forward shock. Thus, the velocity
decrease at the reverse shock is moderate, which results in less thermal
energy creation and no temperature peak. Therefore, the temperature
profile is not consistent with observations.

The value of $\chi^2/\rm dof$ of the wind-K model is close to one
(Table~\ref{tab1}), which means that this model successfully reproduced
the observations. The value of $\chi^2/\rm dof$ of the wind-T model is
almost the same as that of the wind-K model, which means that the
influence of thermal energy is minor.  On the other hand, the values of
$\chi^2/\rm dof$ of the explosion models are much larger than one
(Table~\ref{tab1}), and they are inconsistent with the observations.
 The success of the wind model implies that the duration of the GC
activities ($\sim 10^7$~yr) was not instantaneous. We assumed that
the halo gas was in the CIE state before the wind began blowing. Since
the NEI state returns to the CIE state in a time-scale of
\begin{equation}
 t_{\rm CIE}\sim 10^7\left(\frac{n_{\rm e}}{3\times 10^{-3}\rm\: cm^{-3}}\right)^{-1}\:\rm yr
\end{equation}
\citep{2010ApJ...718..583S,2022MNRAS.512.2034Y}, any wind episodes
before that period do not affect our results.  Our results
($t_0\sim 10^7$~yr) may indicate that the duty cycle of
Galactic centre activity is $\gtrsim 10^7$~yr. We cannot deny the
possibility that a similar episode of energy injection rendered the halo
gas non-equilibrium $\gg 10^7$~yr before.

We note that the actual wind may be multi-phase. In the above
calculations, we implicitly assumed that one phase is dominant over
other phases in the wind gas. However, while the the wind-K and -T
models have different initial temperatures, they give similar results
(Fig.~\ref{fig:wind}). This implies that even if one phase is not
dominant, the overall structure would not change much.

\section{Discussion}
\label{sec:discuss}

The wind power and mass flux obtained above offer clues regarding the
origin of the Fermi bubbles. The best-fit wind-K model indicates that
the energy injection rate is $L_{\rm w}=1.4\times 10^{42}\rm\: erg\:
s^{-1}$. Given the mass of the supermassive black hole in the GC ($\sim
4\times 10^6\: M_\odot$;
\citealp{2016ApJ...830...17B,2017ApJ...837...30G}), $L_{\rm w}$ is much
smaller than the theoretical maximum rate given by the Eddington
limit($\sim 5\times 10^{44}\rm\: erg\: s^{-1}$).  Thus, the black hole
can provide sufficient power to propagate the wind. This is in contrast
to the explosion model, in which the energy input must be instantaneous
and as powerful as the Eddington limit. The best-fit wind model also
indicated that the mass flux is $\dot{M}=3.4\: M_\odot{\rm
yr^{-1}}$. The large flux suggests that the wind is not directly
produced near the black hole but is formed as light energetic outflows
from the black hole entraining the surrounding interstellar gas. For
example, because the black hole in the GC is surrounded by a massive
molecular cloud ($\sim 3\times 10^7\: M_\odot$) called the central
molecular zone \citep{2011ApJ...735L..33M}, a fraction of the cold gas
may be blown out and become the wind. The cold ($\sim 10$--100~K) gas
outflow observed around the GC \citep{2020Natur.584..364D} may be
remnants of the windblown cold gas.  The existence of high-velocity,
low-metallicity clouds may indicate that even clouds in the halo have
been blown away (\citealt{2022arXiv220708838A}, see also
\citealt{2015ApJ...799L...7F,2017ApJ...834..191B,2018ApJ...860...98K,2020ApJ...898..128A}).
In addition, warm ($\sim 10^3$--$10^5$~K) and hot ($\gtrsim 10^6$~K)
interstellar gases around the GC may be accelerated. Notably, outflows
of warm and hot gases have been observed around the GC
\citep{2013ApJ...770L...4M,2015ApJ...799L...7F,2019ApJ...875...32N,2019Natur.567..347P}. These
multiphase gases may have mixed, becoming the wind blowing into the
Fermi bubbles. To entrain a large amount of gas ($\dot{M}=3.4\:
M_\odot{\rm yr^{-1}}$), the initial gas flows generated near the black
hole should have wide opening angles, say $\sim 60$--$100^\circ$, and
large cross-sectional areas. Therefore, they are unlikely to be thin
jets. This type and scale of outflows are often observed around active
galactic nuclei (AGN) in other galaxies
\citep{2013MNRAS.430.1102T,2014MNRAS.443.2154T,2021NatAs...5...13L,2022ApJ...927..176B}.
Thus, the formation of the wind and the process of mass loading around
the GC may be similar to those of AGN outflows
(\citealt{2015ARA&A..53..115K,2021ApJ...922..254C} see also
\citealt{2009AIPC.1201..321S}). The current level of activity of
Sagittarius A$^*$ is much lower than $L_{\rm w}$, suggesting that the
activity ceased only recently or fluctuated on a time scale much shorter
than the wind period of $\sim 10^7$~yr
\citep{1996PASJ...48..249K,2006PASJ...58..965T,2013ApJ...778...58B,2019ApJ...886...45B}.  If the activity
ceases, the mass flux of the wind should decrease near the GC. On the
other hand, the star formation activity and associated supernova
explosions around the GC are unlikely to be the energy sources of the
fast wind and Fermi bubbles. This is because the star formation rate
around the GC has been only $\sim 0.1\: M_\odot\rm yr^{-1}$ for the past
$\gtrsim 10^6$~yr \citep{2009ApJ...702..178Y,2012A&A...537A.121I},
which suggests that the power provided by the activity is only $\sim
10^{40}\rm\: erg\: s^{-1}$ \citep{2015ApJ...808..107C}.

The AGN outflows in other galaxies are thought to regulate the growth of
the galaxies and black holes \citep{2015ARA&A..53..115K}; therefore, it
may be natural that the same phenomenon occurs in the Milky Way
Galaxy. If the wind blowing into the Fermi bubbles is the same as the
AGN outflows, the proximity of the GC can allow us to closely study the
nature of the AGN outflows. Moreover, the temperature and shock age of
the gas associated with the Fermi bubbles will soon be measured more
precisely with the X-Ray Imaging and Spectroscopy Mission (XRISM)
satellite with an exceptional energy resolution
(\citealt{2020SPIE11444E..22T}, see also \citealt{2015PASJ...67...56I}).
The XRISM satellite can also directly measure X-ray gas velocity for the
first time. The velocity gradient in the high-density region ($r\sim
4$--6.5~kpc) is particularly useful because the gradient differs between
the wind and explosion models (Figs.~\ref{fig:wind}
and~\ref{fig:exp}). Observations of the gas velocity around the GC could
also be helpful in understanding the mass loading process of the wind.

Although our results indicate the existence of a strong reverse
shock, prominent non-thermal emissions have not been observed
there. This may imply that the efficency of particle acceleration is
small. One possibility is that magnetic fields on the upstream side of
the shock are turbulent. This may be expected if the winds from the GC
are not laminar. If the magnetic fields are turbulent in the upstream
region, the particles are trapped by the fields and just swept
downstream with the fluids. The particles are not allowed to go back and
forth across the shock. As a result, the mechanism of diffusive shock
acceleration (DSA) does not work \citep{2015ApJ...809...29T}.

Finally, we should discuss caveats on the use of 1D simulations in this
work.  Since the actual Fermi bubbles are two bubbles elongated
perpendicular to the Galactic plane, our 1D model cannot precisely
reproduce the bubbles, especially for $\lesssim 2$~kpc above or below
the Galactic disk, where they are heavily distorted by the dense disk as
is suggested by the gamma-ray images
\citep{2010ApJ...724.1044S,2014ApJ...793...64A}.  However, since
radiative cooling is inefficient, the evolution of the bubbles is rather
simple. Moreover, the impact of the non-spherical gravity on the bubble
evolution is minor as long as the expansion is supersonic. Thus, our
model should reproduce the actual bubbles at least qualitatively. Even
quantitatively, our results are not much different from those of
multi-dimensional simulations, given a bubble volume. For example, the
wind-K model requires total energy of $\sim 4\times 10^{56}$~ergs, which
is consistent with those derived through 3D wind simulations by
\citet{2014ApJ...790..109M} within an order of magnitude. Our value is
larger than those obtained by \citet{2014ApJ...790..109M} probably
because the 1D bubble expands at a low altitude (closer to the Galactic
plane) where the density of the halo gas is larger. The existence of the
wind of $\sim 1000\:\rm km\: s^{-1}$ is robust, because the height of
the temperature peak at the reverse shock simply reflects the upstream
velocity.

\section{Conclusion}
\label{sec:conc}

We have studied the properties of the non-equilibrium X-ray gas
structures around the Fermi bubbles. Using numerical simulations, we
showed that a combination of the density, temperature, and shock age
profiles can be a useful tool to identify the energy injection mechanism
at the GC. By comparing the results of numerical simulations with
observations, we found that the Fermi bubbles were most likely to be
created by a fast wind from the GC, which had a speed of $\sim 1000\rm\:
km\: s^{-1}$, and blew for $\sim 10^7$~yr. The wind generates a strong
reverse shock and reproduces a sharp temperature increase as is
observed.  On the other hand, models in which energy was instantaneously
injected cannot reproduce the observed temperature profile. Considering
the power of the wind, the energy source seems to be Sagittarius A$^*$,
and not star formation activities.  Since the mass flux of the wind is
large, interstellar gas might have been entrained by the wind.  For the
entrainment, the initial outflows launched in the vicinity of the black
hole should have a large opening angle and they are not likely to be
thin jets. The outflows may be the same as AGN outflows that are often
observed in other galaxies.

\section*{Acknowledgements}

We would like to thank the anonymous referee for a constructive report.
Y.F. was supported by JSPS KAKENHI 20H00181, 22H00158, 22H01268.

%%%%%%%%%%%%%%%%%%%%%%%%%%%%%%%%%%%%%%%%%%%%%%%%%%
\section*{Data Availability}

The data generated from computations are reported in the paper and any additional data will be made available upon reasonable request to the corresponding author.



\bibliographystyle{mnras}
\bibliography{fb} % if your bibtex file is called example.bib

\begin{thebibliography}{}
\makeatletter
\relax
\def\mn@urlcharsother{\let\do\@makeother \do\$\do\&\do\#\do\^\do\_\do\%\do\~}
\def\mn@doi{\begingroup\mn@urlcharsother \@ifnextchar [ {\mn@doi@}
  {\mn@doi@[]}}
\def\mn@doi@[#1]#2{\def\@tempa{#1}\ifx\@tempa\@empty \href
  {http://dx.doi.org/#2} {doi:#2}\else \href {http://dx.doi.org/#2} {#1}\fi
  \endgroup}
\def\mn@eprint#1#2{\mn@eprint@#1:#2::\@nil}
\def\mn@eprint@arXiv#1{\href {http://arxiv.org/abs/#1} {{\tt arXiv:#1}}}
\def\mn@eprint@dblp#1{\href {http://dblp.uni-trier.de/rec/bibtex/#1.xml}
  {dblp:#1}}
\def\mn@eprint@#1:#2:#3:#4\@nil{\def\@tempa {#1}\def\@tempb {#2}\def\@tempc
  {#3}\ifx \@tempc \@empty \let \@tempc \@tempb \let \@tempb \@tempa \fi \ifx
  \@tempb \@empty \def\@tempb {arXiv}\fi \@ifundefined
  {mn@eprint@\@tempb}{\@tempb:\@tempc}{\expandafter \expandafter \csname
  mn@eprint@\@tempb\endcsname \expandafter{\@tempc}}}

\bibitem[\protect\citeauthoryear{{Ackermann} et~al.,}{{Ackermann}
  et~al.}{2014}]{2014ApJ...793...64A}
{Ackermann} M.,  et~al., 2014, \mn@doi [\apj] {10.1088/0004-637X/793/1/64},
  \href {https://ui.adsabs.harvard.edu/abs/2014ApJ...793...64A} {793, 64}

\bibitem[\protect\citeauthoryear{{Ashley}, {Fox}, {Jenkins}, {Wakker},
  {Bordoloi}, {Lockman}, {Savage}  \& {Karim}}{{Ashley}
  et~al.}{2020}]{2020ApJ...898..128A}
{Ashley} T.,  {Fox} A.~J.,  {Jenkins} E.~B.,  {Wakker} B.~P.,  {Bordoloi} R.,
  {Lockman} F.~J.,  {Savage} B.~D.,   {Karim} T.,  2020, \mn@doi [\apj]
  {10.3847/1538-4357/ab9ff8}, \href
  {https://ui.adsabs.harvard.edu/abs/2020ApJ...898..128A} {898, 128}

\bibitem[\protect\citeauthoryear{{Ashley}, {Fox}, {Cashman}, {Lockman},
  {Bordoloi}, {Jenkins}, {Wakker}  \& {Karim}}{{Ashley}
  et~al.}{2022}]{2022arXiv220708838A}
{Ashley} T.,  {Fox} A.~J.,  {Cashman} F.~H.,  {Lockman} F.~J.,  {Bordoloi} R.,
  {Jenkins} E.~B.,  {Wakker} B.~P.,   {Karim} T.,  2022, arXiv e-prints, \href
  {https://ui.adsabs.harvard.edu/abs/2022arXiv220708838A} {p. arXiv:2207.08838}

\bibitem[\protect\citeauthoryear{{Bland-Hawthorn} \& {Cohen}}{{Bland-Hawthorn}
  \& {Cohen}}{2003}]{2003ApJ...582..246B}
{Bland-Hawthorn} J.,  {Cohen} M.,  2003, \mn@doi [\apj] {10.1086/344573}, \href
  {https://ui.adsabs.harvard.edu/abs/2003ApJ...582..246B} {582, 246}

\bibitem[\protect\citeauthoryear{{Bland-Hawthorn}, {Maloney}, {Sutherland}  \&
  {Madsen}}{{Bland-Hawthorn} et~al.}{2013}]{2013ApJ...778...58B}
{Bland-Hawthorn} J.,  {Maloney} P.~R.,  {Sutherland} R.~S.,   {Madsen} G.~J.,
  2013, \mn@doi [\apj] {10.1088/0004-637X/778/1/58}, \href
  {https://ui.adsabs.harvard.edu/abs/2013ApJ...778...58B} {778, 58}

\bibitem[\protect\citeauthoryear{{Bland-Hawthorn} et~al.,}{{Bland-Hawthorn}
  et~al.}{2019}]{2019ApJ...886...45B}
{Bland-Hawthorn} J.,  et~al., 2019, \mn@doi [\apj] {10.3847/1538-4357/ab44c8},
  \href {https://ui.adsabs.harvard.edu/abs/2019ApJ...886...45B} {886, 45}

\bibitem[\protect\citeauthoryear{{Bodenheimer}, {Laughlin}, {Rozyczka}  \&
  {Yorke}}{{Bodenheimer} et~al.}{2007}]{2007nmai.conf.....B}
{Bodenheimer} P.,  {Laughlin} G.~P.,  {Rozyczka} M.,   {Yorke} H.~W.,  2007,
  {Numerical Methods in Astrophysics: An Introduction, CRC Press, Boca Raton}

\bibitem[\protect\citeauthoryear{{Boehle} et~al.,}{{Boehle}
  et~al.}{2016}]{2016ApJ...830...17B}
{Boehle} A.,  et~al., 2016, \mn@doi [\apj] {10.3847/0004-637X/830/1/17}, \href
  {https://ui.adsabs.harvard.edu/abs/2016ApJ...830...17B} {830, 17}

\bibitem[\protect\citeauthoryear{{Bohdan}, {Pohl}, {Niemiec}, {Morris},
  {Matsumoto}, {Amano}  \& {Hoshino}}{{Bohdan}
  et~al.}{2020}]{2020ApJ...904...12B}
{Bohdan} A.,  {Pohl} M.,  {Niemiec} J.,  {Morris} P.~J.,  {Matsumoto} Y.,
  {Amano} T.,   {Hoshino} M.,  2020, \mn@doi [\apj] {10.3847/1538-4357/abbc19},
  \href {https://ui.adsabs.harvard.edu/abs/2020ApJ...904...12B} {904, 12}

\bibitem[\protect\citeauthoryear{{Bordoloi} et~al.,}{{Bordoloi}
  et~al.}{2017}]{2017ApJ...834..191B}
{Bordoloi} R.,  et~al., 2017, \mn@doi [\apj] {10.3847/1538-4357/834/2/191},
  \href {https://ui.adsabs.harvard.edu/abs/2017ApJ...834..191B} {834, 191}

\bibitem[\protect\citeauthoryear{{Byun}, {Arav}  \& {Hall}}{{Byun}
  et~al.}{2022}]{2022ApJ...927..176B}
{Byun} D.,  {Arav} N.,   {Hall} P.~B.,  2022, \mn@doi [\apj]
  {10.3847/1538-4357/ac503d}, \href
  {https://ui.adsabs.harvard.edu/abs/2022ApJ...927..176B} {927, 176}

\bibitem[\protect\citeauthoryear{{Carretti} et~al.,}{{Carretti}
  et~al.}{2013}]{2013Natur.493...66C}
{Carretti} E.,  et~al., 2013, \mn@doi [\nat] {10.1038/nature11734}, \href
  {https://ui.adsabs.harvard.edu/abs/2013Natur.493...66C} {493, 66}

\bibitem[\protect\citeauthoryear{{Cecil}, {Wagner}, {Bland-Hawthorn},
  {Bicknell}  \& {Mukherjee}}{{Cecil} et~al.}{2021}]{2021ApJ...922..254C}
{Cecil} G.,  {Wagner} A.~Y.,  {Bland-Hawthorn} J.,  {Bicknell} G.~V.,
  {Mukherjee} D.,  2021, \mn@doi [\apj] {10.3847/1538-4357/ac224f}, \href
  {https://ui.adsabs.harvard.edu/abs/2021ApJ...922..254C} {922, 254}

\bibitem[\protect\citeauthoryear{{Crocker}}{{Crocker}}{2012}]{2012MNRAS.423.3512C}
{Crocker} R.~M.,  2012, \mn@doi [\mnras] {10.1111/j.1365-2966.2012.21149.x},
  \href {https://ui.adsabs.harvard.edu/abs/2012MNRAS.423.3512C} {423, 3512}

\bibitem[\protect\citeauthoryear{{Crocker} \& {Aharonian}}{{Crocker} \&
  {Aharonian}}{2011}]{2011PhRvL.106j1102C}
{Crocker} R.~M.,  {Aharonian} F.,  2011, \mn@doi [\prl]
  {10.1103/PhysRevLett.106.101102}, \href
  {https://ui.adsabs.harvard.edu/abs/2011PhRvL.106j1102C} {106, 101102}

\bibitem[\protect\citeauthoryear{{Crocker}, {Bicknell}, {Taylor}  \&
  {Carretti}}{{Crocker} et~al.}{2015}]{2015ApJ...808..107C}
{Crocker} R.~M.,  {Bicknell} G.~V.,  {Taylor} A.~M.,   {Carretti} E.,  2015,
  \mn@doi [\apj] {10.1088/0004-637X/808/2/107}, \href
  {https://ui.adsabs.harvard.edu/abs/2015ApJ...808..107C} {808, 107}

\bibitem[\protect\citeauthoryear{{Di Teodoro}, {McClure-Griffiths}, {Lockman}
  \& {Armillotta}}{{Di Teodoro} et~al.}{2020}]{2020Natur.584..364D}
{Di Teodoro} E.~M.,  {McClure-Griffiths} N.~M.,  {Lockman} F.~J.,
  {Armillotta} L.,  2020, \mn@doi [\nat] {10.1038/s41586-020-2595-z}, \href
  {https://ui.adsabs.harvard.edu/abs/2020Natur.584..364D} {584, 364}

\bibitem[\protect\citeauthoryear{{Dobler} \& {Finkbeiner}}{{Dobler} \&
  {Finkbeiner}}{2008}]{2008ApJ...680.1222D}
{Dobler} G.,  {Finkbeiner} D.~P.,  2008, \mn@doi [\apj] {10.1086/587862}, \href
  {https://ui.adsabs.harvard.edu/abs/2008ApJ...680.1222D} {680, 1222}

\bibitem[\protect\citeauthoryear{{Dobler}, {Finkbeiner}, {Cholis}, {Slatyer}
  \& {Weiner}}{{Dobler} et~al.}{2010}]{2010ApJ...717..825D}
{Dobler} G.,  {Finkbeiner} D.~P.,  {Cholis} I.,  {Slatyer} T.,   {Weiner} N.,
  2010, \mn@doi [\apj] {10.1088/0004-637X/717/2/825}, \href
  {https://ui.adsabs.harvard.edu/abs/2010ApJ...717..825D} {717, 825}

\bibitem[\protect\citeauthoryear{{Finkbeiner}}{{Finkbeiner}}{2004}]{2004ApJ...614..186F}
{Finkbeiner} D.~P.,  2004, \mn@doi [\apj] {10.1086/423482}, \href
  {https://ui.adsabs.harvard.edu/abs/2004ApJ...614..186F} {614, 186}

\bibitem[\protect\citeauthoryear{{Fox} et~al.,}{{Fox}
  et~al.}{2015}]{2015ApJ...799L...7F}
{Fox} A.~J.,  et~al., 2015, \mn@doi [\apjl] {10.1088/2041-8205/799/1/L7}, \href
  {https://ui.adsabs.harvard.edu/abs/2015ApJ...799L...7F} {799, L7}

\bibitem[\protect\citeauthoryear{{Fujita} et~al.,}{{Fujita}
  et~al.}{2008}]{2008PASJ...60.1133F}
{Fujita} Y.,  et~al., 2008, \mn@doi [\pasj] {10.1093/pasj/60.5.1133}, \href
  {https://ui.adsabs.harvard.edu/abs/2008PASJ...60.1133F} {60, 1133}

\bibitem[\protect\citeauthoryear{{Fujita}, {Ohira}  \& {Yamazaki}}{{Fujita}
  et~al.}{2013}]{2013ApJ...775L..20F}
{Fujita} Y.,  {Ohira} Y.,   {Yamazaki} R.,  2013, \mn@doi [\apjl]
  {10.1088/2041-8205/775/1/L20}, \href
  {https://ui.adsabs.harvard.edu/abs/2013ApJ...775L..20F} {775, L20}

\bibitem[\protect\citeauthoryear{{Fujita}, {Ohira}  \& {Yamazaki}}{{Fujita}
  et~al.}{2014}]{2014ApJ...789...67F}
{Fujita} Y.,  {Ohira} Y.,   {Yamazaki} R.,  2014, \mn@doi [\apj]
  {10.1088/0004-637X/789/1/67}, \href
  {https://ui.adsabs.harvard.edu/abs/2014ApJ...789...67F} {789, 67}

\bibitem[\protect\citeauthoryear{{Gillessen} et~al.,}{{Gillessen}
  et~al.}{2017}]{2017ApJ...837...30G}
{Gillessen} S.,  et~al., 2017, \mn@doi [\apj] {10.3847/1538-4357/aa5c41}, \href
  {https://ui.adsabs.harvard.edu/abs/2017ApJ...837...30G} {837, 30}

\bibitem[\protect\citeauthoryear{{Guo} \& {Mathews}}{{Guo} \&
  {Mathews}}{2012}]{2012ApJ...756..181G}
{Guo} F.,  {Mathews} W.~G.,  2012, \mn@doi [\apj]
  {10.1088/0004-637X/756/2/181}, \href
  {https://ui.adsabs.harvard.edu/abs/2012ApJ...756..181G} {756, 181}

\bibitem[\protect\citeauthoryear{{Immer}, {Schuller}, {Omont}  \&
  {Menten}}{{Immer} et~al.}{2012}]{2012A&A...537A.121I}
{Immer} K.,  {Schuller} F.,  {Omont} A.,   {Menten} K.~M.,  2012, \mn@doi
  [\aap] {10.1051/0004-6361/201117857}, \href
  {https://ui.adsabs.harvard.edu/abs/2012A&A...537A.121I} {537, A121}

\bibitem[\protect\citeauthoryear{{Inoue}, {Nakashima}, {Tahara}, {Kataoka},
  {Totani}, {Fujita}  \& {Sofue}}{{Inoue} et~al.}{2015}]{2015PASJ...67...56I}
{Inoue} Y.,  {Nakashima} S.,  {Tahara} M.,  {Kataoka} J.,  {Totani} T.,
  {Fujita} Y.,   {Sofue} Y.,  2015, \mn@doi [\pasj] {10.1093/pasj/psv032},
  \href {https://ui.adsabs.harvard.edu/abs/2015PASJ...67...56I} {67, 56}

\bibitem[\protect\citeauthoryear{{Karim} et~al.,}{{Karim}
  et~al.}{2018}]{2018ApJ...860...98K}
{Karim} T.,  et~al., 2018, \mn@doi [\apj] {10.3847/1538-4357/aac167}, \href
  {https://ui.adsabs.harvard.edu/abs/2018ApJ...860...98K} {860, 98}

\bibitem[\protect\citeauthoryear{{Kataoka} et~al.,}{{Kataoka}
  et~al.}{2013}]{2013ApJ...779...57K}
{Kataoka} J.,  et~al., 2013, \mn@doi [\apj] {10.1088/0004-637X/779/1/57}, \href
  {https://ui.adsabs.harvard.edu/abs/2013ApJ...779...57K} {779, 57}

\bibitem[\protect\citeauthoryear{{Keshet} \& {Gurwich}}{{Keshet} \&
  {Gurwich}}{2018}]{2018MNRAS.480..223K}
{Keshet} U.,  {Gurwich} I.,  2018, \mn@doi [\mnras] {10.1093/mnras/sty1533},
  \href {https://ui.adsabs.harvard.edu/abs/2018MNRAS.480..223K} {480, 223}

\bibitem[\protect\citeauthoryear{{King} \& {Pounds}}{{King} \&
  {Pounds}}{2015}]{2015ARA&A..53..115K}
{King} A.,  {Pounds} K.,  2015, \mn@doi [\araa]
  {10.1146/annurev-astro-082214-122316}, \href
  {https://ui.adsabs.harvard.edu/abs/2015ARA&A..53..115K} {53, 115}

\bibitem[\protect\citeauthoryear{{Koyama}, {Maeda}, {Sonobe}, {Takeshima},
  {Tanaka}  \& {Yamauchi}}{{Koyama} et~al.}{1996}]{1996PASJ...48..249K}
{Koyama} K.,  {Maeda} Y.,  {Sonobe} T.,  {Takeshima} T.,  {Tanaka} Y.,
  {Yamauchi} S.,  1996, \mn@doi [\pasj] {10.1093/pasj/48.2.249}, \href
  {https://ui.adsabs.harvard.edu/abs/1996PASJ...48..249K} {48, 249}

\bibitem[\protect\citeauthoryear{{LaRocca}, {Kaaret}, {Kuntz}, {Hodges-Kluck},
  {Zajczyk}, {Bluem}, {Ringuette}  \& {Jahoda}}{{LaRocca}
  et~al.}{2020}]{2020ApJ...904...54L}
{LaRocca} D.~M.,  {Kaaret} P.,  {Kuntz} K.~D.,  {Hodges-Kluck} E.,  {Zajczyk}
  A.,  {Bluem} J.,  {Ringuette} R.,   {Jahoda} K.~M.,  2020, \mn@doi [\apj]
  {10.3847/1538-4357/abbdfd}, \href
  {https://ui.adsabs.harvard.edu/abs/2020ApJ...904...54L} {904, 54}

\bibitem[\protect\citeauthoryear{{Lacki}}{{Lacki}}{2014}]{2014MNRAS.444L..39L}
{Lacki} B.~C.,  2014, \mn@doi [\mnras] {10.1093/mnrasl/slu107}, \href
  {https://ui.adsabs.harvard.edu/abs/2014MNRAS.444L..39L} {444, L39}

\bibitem[\protect\citeauthoryear{{Laha}, {Reynolds}, {Reeves}, {Kriss},
  {Guainazzi}, {Smith}, {Veilleux}  \& {Proga}}{{Laha}
  et~al.}{2021}]{2021NatAs...5...13L}
{Laha} S.,  {Reynolds} C.~S.,  {Reeves} J.,  {Kriss} G.,  {Guainazzi} M.,
  {Smith} R.,  {Veilleux} S.,   {Proga} D.,  2021, \mn@doi [Nature Astronomy]
  {10.1038/s41550-020-01255-2}, \href
  {https://ui.adsabs.harvard.edu/abs/2021NatAs...5...13L} {5, 13}

\bibitem[\protect\citeauthoryear{{McClure-Griffiths}, {Green}, {Hill},
  {Lockman}, {Dickey}, {Gaensler}  \& {Green}}{{McClure-Griffiths}
  et~al.}{2013}]{2013ApJ...770L...4M}
{McClure-Griffiths} N.~M.,  {Green} J.~A.,  {Hill} A.~S.,  {Lockman} F.~J.,
  {Dickey} J.~M.,  {Gaensler} B.~M.,   {Green} A.~J.,  2013, \mn@doi [\apjl]
  {10.1088/2041-8205/770/1/L4}, \href
  {https://ui.adsabs.harvard.edu/abs/2013ApJ...770L...4M} {770, L4}

\bibitem[\protect\citeauthoryear{{Miyamoto} \& {Nagai}}{{Miyamoto} \&
  {Nagai}}{1975}]{1975PASJ...27..533M}
{Miyamoto} M.,  {Nagai} R.,  1975, \pasj, \href
  {https://ui.adsabs.harvard.edu/abs/1975PASJ...27..533M} {27, 533}

\bibitem[\protect\citeauthoryear{{Molinari} et~al.,}{{Molinari}
  et~al.}{2011}]{2011ApJ...735L..33M}
{Molinari} S.,  et~al., 2011, \mn@doi [\apjl] {10.1088/2041-8205/735/2/L33},
  \href {https://ui.adsabs.harvard.edu/abs/2011ApJ...735L..33M} {735, L33}

\bibitem[\protect\citeauthoryear{{Mondal}, {Keshet}, {Sarkar}  \&
  {Gurwich}}{{Mondal} et~al.}{2022}]{2022MNRAS.514.2581M}
{Mondal} S.,  {Keshet} U.,  {Sarkar} K.~C.,   {Gurwich} I.,  2022, \mn@doi
  [\mnras] {10.1093/mnras/stac1084}, \href
  {https://ui.adsabs.harvard.edu/abs/2022MNRAS.514.2581M} {514, 2581}

\bibitem[\protect\citeauthoryear{{Mou}, {Yuan}, {Bu}, {Sun}  \& {Su}}{{Mou}
  et~al.}{2014}]{2014ApJ...790..109M}
{Mou} G.,  {Yuan} F.,  {Bu} D.,  {Sun} M.,   {Su} M.,  2014, \mn@doi [\apj]
  {10.1088/0004-637X/790/2/109}, \href
  {https://ui.adsabs.harvard.edu/abs/2014ApJ...790..109M} {790, 109}

\bibitem[\protect\citeauthoryear{{Mou}, {Yuan}, {Gan}  \& {Sun}}{{Mou}
  et~al.}{2015}]{2015ApJ...811...37M}
{Mou} G.,  {Yuan} F.,  {Gan} Z.,   {Sun} M.,  2015, \mn@doi [\apj]
  {10.1088/0004-637X/811/1/37}, \href
  {https://ui.adsabs.harvard.edu/abs/2015ApJ...811...37M} {811, 37}

\bibitem[\protect\citeauthoryear{{Nakashima}, {Koyama}, {Wang}  \&
  {Enokiya}}{{Nakashima} et~al.}{2019}]{2019ApJ...875...32N}
{Nakashima} S.,  {Koyama} K.,  {Wang} Q.~D.,   {Enokiya} R.,  2019, \mn@doi
  [\apj] {10.3847/1538-4357/ab0d82}, \href
  {https://ui.adsabs.harvard.edu/abs/2019ApJ...875...32N} {875, 32}

\bibitem[\protect\citeauthoryear{{Narayanan} \& {Slatyer}}{{Narayanan} \&
  {Slatyer}}{2017}]{2017MNRAS.468.3051N}
{Narayanan} S.~A.,  {Slatyer} T.~R.,  2017, \mn@doi [\mnras]
  {10.1093/mnras/stx577}, \href
  {https://ui.adsabs.harvard.edu/abs/2017MNRAS.468.3051N} {468, 3051}

\bibitem[\protect\citeauthoryear{{Planck Collaboration}}{{Planck
  Collaboration}}{2013}]{2013A&A...554A.139P}
{Planck Collaboration} 2013, \mn@doi [\aap] {10.1051/0004-6361/201220271},
  \href {https://ui.adsabs.harvard.edu/abs/2013A&A...554A.139P} {554, A139}

\bibitem[\protect\citeauthoryear{{Ponti} et~al.,}{{Ponti}
  et~al.}{2019}]{2019Natur.567..347P}
{Ponti} G.,  et~al., 2019, \mn@doi [\nat] {10.1038/s41586-019-1009-6}, \href
  {https://ui.adsabs.harvard.edu/abs/2019Natur.567..347P} {567, 347}

\bibitem[\protect\citeauthoryear{{Predehl} et~al.,}{{Predehl}
  et~al.}{2020}]{2020Natur.588..227P}
{Predehl} P.,  et~al., 2020, \mn@doi [\nat] {10.1038/s41586-020-2979-0}, \href
  {https://ui.adsabs.harvard.edu/abs/2020Natur.588..227P} {588, 227}

\bibitem[\protect\citeauthoryear{{Safi-Harb}, {Harrus}, {Petre}, {Pavlov},
  {Koptsevich}  \& {Sanwal}}{{Safi-Harb} et~al.}{2001}]{2001ApJ...561..308S}
{Safi-Harb} S.,  {Harrus} I.~M.,  {Petre} R.,  {Pavlov} G.~G.,  {Koptsevich}
  A.~B.,   {Sanwal} D.,  2001, \mn@doi [\apj] {10.1086/322978}, \href
  {https://ui.adsabs.harvard.edu/abs/2001ApJ...561..308S} {561, 308}

\bibitem[\protect\citeauthoryear{{Salter}}{{Salter}}{1983}]{1983BASI...11....1S}
{Salter} C.~J.,  1983, Bulletin of the Astronomical Society of India, \href
  {https://ui.adsabs.harvard.edu/abs/1983BASI...11....1S} {11, 1}

\bibitem[\protect\citeauthoryear{{Sarkar}, {Nath}  \& {Sharma}}{{Sarkar}
  et~al.}{2015}]{2015MNRAS.453.3827S}
{Sarkar} K.~C.,  {Nath} B.~B.,   {Sharma} P.,  2015, \mn@doi [\mnras]
  {10.1093/mnras/stv1806}, \href
  {https://ui.adsabs.harvard.edu/abs/2015MNRAS.453.3827S} {453, 3827}

\bibitem[\protect\citeauthoryear{{Smith} \& {Hughes}}{{Smith} \&
  {Hughes}}{2010}]{2010ApJ...718..583S}
{Smith} R.~K.,  {Hughes} J.~P.,  2010, \mn@doi [\apj]
  {10.1088/0004-637X/718/1/583}, \href
  {https://ui.adsabs.harvard.edu/abs/2010ApJ...718..583S} {718, 583}

\bibitem[\protect\citeauthoryear{{Snowden} et~al.,}{{Snowden}
  et~al.}{1995}]{1995ApJ...454..643S}
{Snowden} S.~L.,  et~al., 1995, \mn@doi [\apj] {10.1086/176517}, \href
  {https://ui.adsabs.harvard.edu/abs/1995ApJ...454..643S} {454, 643}

\bibitem[\protect\citeauthoryear{{Sofue}}{{Sofue}}{2000}]{2000ApJ...540..224S}
{Sofue} Y.,  2000, \mn@doi [\apj] {10.1086/309297}, \href
  {https://ui.adsabs.harvard.edu/abs/2000ApJ...540..224S} {540, 224}

\bibitem[\protect\citeauthoryear{{Soker}, {Sternberg}  \& {Pizzolato}}{{Soker}
  et~al.}{2009}]{2009AIPC.1201..321S}
{Soker} N.,  {Sternberg} A.,   {Pizzolato} F.,  2009, in {Heinz} S.,  {Wilcots}
  E.,  eds,  American Institute of Physics Conference Series Vol. 1201, The
  Monster's Fiery Breath: Feedback in Galaxies, Groups, and Clusters. pp
  321--325 (\mn@eprint {arXiv} {0909.0220}), \mn@doi{10.1063/1.3293066}

\bibitem[\protect\citeauthoryear{{Su}, {Slatyer}  \& {Finkbeiner}}{{Su}
  et~al.}{2010}]{2010ApJ...724.1044S}
{Su} M.,  {Slatyer} T.~R.,   {Finkbeiner} D.~P.,  2010, \mn@doi [\apj]
  {10.1088/0004-637X/724/2/1044}, \href
  {https://ui.adsabs.harvard.edu/abs/2010ApJ...724.1044S} {724, 1044}

\bibitem[\protect\citeauthoryear{{Takamoto} \& {Kirk}}{{Takamoto} \&
  {Kirk}}{2015}]{2015ApJ...809...29T}
{Takamoto} M.,  {Kirk} J.~G.,  2015, \mn@doi [\apj]
  {10.1088/0004-637X/809/1/29}, \href
  {https://ui.adsabs.harvard.edu/abs/2015ApJ...809...29T} {809, 29}

\bibitem[\protect\citeauthoryear{{Tashiro} et~al.,}{{Tashiro}
  et~al.}{2020}]{2020SPIE11444E..22T}
{Tashiro} M.,  et~al., 2020, in Society of Photo-Optical Instrumentation
  Engineers (SPIE) Conference Series. p. 1144422, \mn@doi{10.1117/12.2565812}

\bibitem[\protect\citeauthoryear{{Tombesi}, {Cappi}, {Reeves}, {Nemmen},
  {Braito}, {Gaspari}  \& {Reynolds}}{{Tombesi}
  et~al.}{2013}]{2013MNRAS.430.1102T}
{Tombesi} F.,  {Cappi} M.,  {Reeves} J.~N.,  {Nemmen} R.~S.,  {Braito} V.,
  {Gaspari} M.,   {Reynolds} C.~S.,  2013, \mn@doi [\mnras]
  {10.1093/mnras/sts692}, \href
  {https://ui.adsabs.harvard.edu/abs/2013MNRAS.430.1102T} {430, 1102}

\bibitem[\protect\citeauthoryear{{Tombesi}, {Tazaki}, {Mushotzky}, {Ueda},
  {Cappi}, {Gofford}, {Reeves}  \& {Guainazzi}}{{Tombesi}
  et~al.}{2014}]{2014MNRAS.443.2154T}
{Tombesi} F.,  {Tazaki} F.,  {Mushotzky} R.~F.,  {Ueda} Y.,  {Cappi} M.,
  {Gofford} J.,  {Reeves} J.~N.,   {Guainazzi} M.,  2014, \mn@doi [\mnras]
  {10.1093/mnras/stu1297}, \href
  {https://ui.adsabs.harvard.edu/abs/2014MNRAS.443.2154T} {443, 2154}

\bibitem[\protect\citeauthoryear{{Totani}}{{Totani}}{2006}]{2006PASJ...58..965T}
{Totani} T.,  2006, \mn@doi [\pasj] {10.1093/pasj/58.6.965}, \href
  {https://ui.adsabs.harvard.edu/abs/2006PASJ...58..965T} {58, 965}

\bibitem[\protect\citeauthoryear{{Yamamoto}, {Kataoka}  \& {Sofue}}{{Yamamoto}
  et~al.}{2022}]{2022MNRAS.512.2034Y}
{Yamamoto} M.,  {Kataoka} J.,   {Sofue} Y.,  2022, \mn@doi [\mnras]
  {10.1093/mnras/stac577}, \href
  {https://ui.adsabs.harvard.edu/abs/2022MNRAS.512.2034Y} {512, 2034}

\bibitem[\protect\citeauthoryear{{Yang}, {Ruszkowski}, {Ricker}, {Zweibel}  \&
  {Lee}}{{Yang} et~al.}{2012}]{2012ApJ...761..185Y}
{Yang} H. Y.~K.,  {Ruszkowski} M.,  {Ricker} P.~M.,  {Zweibel} E.,   {Lee} D.,
  2012, \mn@doi [\apj] {10.1088/0004-637X/761/2/185}, \href
  {https://ui.adsabs.harvard.edu/abs/2012ApJ...761..185Y} {761, 185}

\bibitem[\protect\citeauthoryear{{Yang}, {Ruszkowski}  \& {Zweibel}}{{Yang}
  et~al.}{2022}]{2022NatAs...6..584Y}
{Yang} H. Y.~K.,  {Ruszkowski} M.,   {Zweibel} E.~G.,  2022, \mn@doi [Nature
  Astronomy] {10.1038/s41550-022-01618-x}, \href
  {https://ui.adsabs.harvard.edu/abs/2022NatAs...6..584Y} {6, 584}

\bibitem[\protect\citeauthoryear{{Yoshino} et~al.,}{{Yoshino}
  et~al.}{2009}]{2009PASJ...61..805Y}
{Yoshino} T.,  et~al., 2009, \mn@doi [\pasj] {10.1093/pasj/61.4.805}, \href
  {https://ui.adsabs.harvard.edu/abs/2009PASJ...61..805Y} {61, 805}

\bibitem[\protect\citeauthoryear{{Yusef-Zadeh} et~al.,}{{Yusef-Zadeh}
  et~al.}{2009}]{2009ApJ...702..178Y}
{Yusef-Zadeh} F.,  et~al., 2009, \mn@doi [\apj] {10.1088/0004-637X/702/1/178},
  \href {https://ui.adsabs.harvard.edu/abs/2009ApJ...702..178Y} {702, 178}

\bibitem[\protect\citeauthoryear{{Zhang} \& {Guo}}{{Zhang} \&
  {Guo}}{2020}]{2020ApJ...894..117Z}
{Zhang} R.,  {Guo} F.,  2020, \mn@doi [\apj] {10.3847/1538-4357/ab8bd0}, \href
  {https://ui.adsabs.harvard.edu/abs/2020ApJ...894..117Z} {894, 117}

\bibitem[\protect\citeauthoryear{{Zubovas} \& {Nayakshin}}{{Zubovas} \&
  {Nayakshin}}{2012}]{2012MNRAS.424..666Z}
{Zubovas} K.,  {Nayakshin} S.,  2012, \mn@doi [\mnras]
  {10.1111/j.1365-2966.2012.21250.x}, \href
  {https://ui.adsabs.harvard.edu/abs/2012MNRAS.424..666Z} {424, 666}

\makeatother
\end{thebibliography}

%\begin{thebibliography}{99}
%\end{thebibliography}

%%%%%%%%%%%%%%%%%%%%%%%%%%%%%%%%%%%%%%%%%%%%%%%%%%

%%%%%%%%%%%%%%%%% APPENDICES %%%%%%%%%%%%%%%%%%%%%

%\appendix

%%%%%%%%%%%%%%%%%%%%%%%%%%%%%%%%%%%%%%%%%%%%%%%%%%

% Don't change these lines
\bsp	% typesetting comment
\label{lastpage}
\end{document}